\documentclass[twocolumn,showpacs,preprintnumbers,amsmath,amssymb]{revtex4}
\usepackage{graphicx}
\usepackage{dcolumn}
\usepackage{bm}

\begin{document}

\newcommand{\be}{\begin{eqnarray}}
\newcommand{\ee}{\end{eqnarray}}
\newcommand{\nn}{\nonumber\\}
\newcommand{\nin}{\noindent}
\newcommand{\la}{\langle}
\newcommand{\ra}{\rangle}

\renewcommand{\theequation}{\arabic{section}.\arabic{equation}}

\title{Magnetic order, Bose-Einstein condensation, and superfluidity\\
 in a bosonic 
$t$-$J$ model of  CP$^1$ spinons and doped Higgs holons}

\author{Koji Aoki$^\ast$} 
\author{Kazuhiko Sakakibara$^\star$}
\author{Ikuo Ichinose$^\ast$}
\author{Tetsuo Matsui$^\dag$}
 \affiliation{${}^\ast$Department of Applied Physics, Graduate School of 
Engineering, \\
Nagoya Institute of Technology, 
Nagoya, 466-8555 Japan 
}
\affiliation{%
${}^\star$Department of Physics, Nara National College of Technology, 
Yamatokohriyama, 639-1080 Japan
}%
\affiliation{%
${}^\dag$Department of Physics, Kinki University, 
Higashi-Osaka, 577-8502 Japan
}%

\date{\today}

\begin{abstract}
We study the three-dimensional U(1) lattice gauge 
theory of a CP$^1$ spinon (Schwinger boson) field and a  Higgs field.
It is a bosonic $t$-$J$ model in slave-particle representation, 
describing the antiferromagnetic (AF) 
Heisenberg spin model with doped bosonic holes expressed by the Higgs field.
The spinon coupling term of the action  favors AF long-range order, whereas
the holon hopping term in the ferromagnetic channel favors 
Bose-Einstein condensation (BEC) of holons.
We investigate the phase structure by means of Monte-Carlo 
simulations and study an interplay of AF order and BEC of holes.
We consider the two variations of the model; (i)
the three-dimensional model at finite temperatures, and
(ii) the two-dimensional model at vanishing temperature.
In the model (i) we find that the AF order and BEC coexist at low temperatures
and certain hole concentrations.
In the model (ii), by varying the hole concentration and the stiffness 
of AF spin coupling, we find a phase diagram similar to the model (i).
Implications of the results to  systems 
of cold atoms and the fermionic $t$-$J$ model of strongly-correlated electrons
are discussed.

\end{abstract}

\maketitle

\section{Introduction}
\setcounter{equation}{0} 

In recent years,  
 cold atoms have attracted interest
of many condensed-matter physicists.
Systems of cold atoms have exhibited (and shall exhibit) various interesting properties
like Bose-Einstein condensation (BEC), superfluidity (SF) caused by the BEC,
 magnetic ordering, etc.
Both for condensed-matter experimentalists and theorists cold atoms offer
an ideal testing ground to develop and check their ideas 
because one can precisely control parameters characterizing these systems like
dimensionality of the system, strength of interaction among atoms, 
concentration of atoms, etc. 
An example of  these ideas is a possible interplay of 
magnetic ordering and BEC (or equivalently SF), 
although coexistence of these two
orders seems not to have been reported experimentally so far.

A standard model of cold bosonic atoms with repulsive interactions
may be the bosonic Hubbard model in which
electrons are replaced by bosonic atoms. 
For definiteness, one may consider hard-core bosons to describe these bosons.
Then, from such a bosonic Hubbard model one may
derive the {\it bosonic} $t$-$J$ model as its effective model
for the case of  strong on-site repulsion and small hole concentrations.
This derivation is achieved just by following the steps 
developed in the theory of high-$T_{\rm c}$ superconductivity
to derive the fermionic $t$-$J$ model from
the standard Hubbard model by tracing out the double-occupancy states.

Therefore, to study the interplay of magnetic ordering and BEC
of bosonic atoms,
it is natural to start from the bosonic $t$-$J$ model.
In fact, very recently, Boninsegni and Prokof'ev\cite{btj}
studied this phenomenon by using the bosonic $t$-$J$ model
where each bosonic electron is considered as a hard-core boson. 
They applied this model to study  bosonic cold 
atoms\cite{cold_boson, cold}.
Because the model is purely bosonic,
one can employ numerical analysis.
By quantum Monte Carlo(MC) simulations, they
studied the low-temperature ($T$) phase diagram
of the two-dimensional (2D) model
for the case of 
{\it anisotropic spin coupling} $J_{x,y}=\alpha J_z, \alpha < 1,$
and found the coexistence region of AF order and BEC
as a result of the phase separation of hole-free and hole-rich
phases. 

The bosonic $t$-$J$ model has another important reason to study,
i.e., it resembles to the fermionic $t$-$J$ model of the 
high-temperature superconductors. 
There the interplay of magnetism and superconductivity(SC) 
is of interest as one of the
most interesting problems in strongly-correlated electron systems.
At present, it is known that SC and antiferromagnetic (AF) N\'eel order
can coexist in clean and uniform samples of the high-$T_c$ cuprates
\cite{mukuda}. Some theoretical works also 
report the coexistence of SC and AF order at $T=0$\cite{sorella}. 
As this phenomenon appears as a result of interplay of fluctuations of 
quantum spins and BEC of superconducting pairs,
simple mean-field-like approximations are inadequate to obtain reliable
results to the relevant questions, e.g., whether  
the $t$-$J$ model of electrons exhibits this coexisting phase.

Not only these AF and SC transitions, the fermionic $t$-$J$ model
is expected to describe the metal-insulator transition(MIT)
as observed in the cuprate high-$T_{\rm c}$ superconductors\cite{mukuda}.
At present, it seems that no well-accepted
 theoretical accounts for MIT beyond the mean field theory
have appeared. 

In these situation, study of the {\it bosonic} $t$-$J$ model
beyond the mean-field theory shall certainly 
shed some lights for understanding of these interesting
problems in the fermionic $t$-$J$ model.
Then one can take advantage of the bosonic nature of 
involved variables, which  affords us 
to perform direct numerical simulations.

In this paper, we study the bosonic $t$-$J$ model.
As explained above, our main motivations are the
following two points; \\
\noindent
(a) studying the interplay of AF and BEC in the cold atoms;\\
\noindent
(b) getting  insight for the AF, MIT and SC 
transitions of cuprate superconductors. \\
\noindent
We shall introduce a new 
representation of the bosonic $t$-$J$ model for 
a $s={1 \over 2}$ 
isotropic AF magnet with doped bosonic holes, 
and investigate the phase structure of the model 
by means of the MC simulations.
Explicitly, we start with {\it the slave-fermion representation}
of the original $t$-$J$ model of electrons 
where the spinons are described by a CP$^1$
(complex projective) field and holons are
described by a one-component fermion field.
Then we replace the fermion field by a Higgs field
with fixed amplitude (U(1) phase variables) to obtain the 
bosonic $t$-$J$ model.
Thus the present model can be regarded as a
{\it slave-particle representation} of the 
bosonic $t$-$J$ model 
that is a canonical model for  cold bosonic atoms in optical lattices.

The usefulness of the slave-particle representation 
in various aspects has been
pointed out  for the original fermionic 
$t$-$J$ model. We expect that similar advantage
of the slave-particle picture holds also
in the  bosonic $t$-$J$ model.
One example is given
by a recent paper by Kaul et al.\cite{sachdev}.
They argue a close resemblance in the phase structure between the
fermionic $t$-$J$ model and the bosonic one.
Gapless fermions with finite density should 
destroy the N\'eel order. Furthermore, they  induce
a phenomenon similar to the Anderson-Higgs mechanism 
to the U(1) gauge dynamics, i.e., they suppress fluctuations of 
the gauge field strongly 
and the gauge dynamics is realized in a deconfinement phase\cite{IMO}.
Similar phenomenon is known to occur in the massless Schwinger model,
i.e., 
$(1+1)$ dimensional quantum electrodynamics, in which the long-range Coulomb
interaction is screened by the gapless ``electron" with a finite
density of states\cite{Schwinger}.
The  Higgs field introduced in the present model to describe bosonic holons 
plays a  role similar to these fermions, and 
then study of the bosonic $t$-$J$ model may give important insight to
the fermionic $t$-$J$ model, in particular, properties of low-energy 
excitations in each phase.

The present paper is organized as follows.
In Sec.2, we shall introduce the bosonic $t$-$J$ model in the
CP$^1$-spinon and Higgs-holon representation.
We first  consider the three-dimensional(3D) model at
finite-$T$.
An effective action is obtained directly 
from the Hamiltonian of the fermionic
$t$-$J$ model by replacing the fermionic holon by bosonic Higgs holon.
In Sec.3, we exhibit the results of the numerical study of the
model and phase diagram of the model.
We calculated the specific heat, the spin and electron-pair
correlation functions,
and monopole density.
From these results,
we conclude that there exists a coexisting phase of AF long-range order and
the BEC in a region of low-$T$ and intermediate hole concentration.
In Sec.4, we shall consider the two-dimensional(2D) system
at vanishing $T$.
We briefly review the derivation of the model.
In Sec.5 we exhibits various results of the numerical calculations and 
the phase diagram.
Section 6 is devoted for conclusion.

\section{Bosonic $t$-$J$ model in the 
CP$^1$-Higgs representation:
3D model at finite $T$'s}
\setcounter{equation}{0} 

In this section, we shall introduce the bosonic $t$-$J$ model
in the CP$^1$ spinon representation.
In particular, we first focus on its phase structure at finite 
temperature($T$).
To be explicit, let us start with the spatially
three-dimensional (3D) original $t$-$J$ model whose 
Hamiltonian $H$ is given by
\be
H&=&-t\sum_{x,\mu,\sigma}\big(\tilde{C}^\dagger_{x+\mu,\sigma}
\tilde{C}_{x\sigma}+{\rm H.c.}\big)
+J\sum_{x,\mu}\vec{\hat{S}}_{x+\mu}\cdot\vec{\hat{S}}_x,\nn
\tilde{C}_{x\sigma} &\equiv& (1-C^\dagger_{x\bar{\sigma}}
C_{x\bar{\sigma}})\ C_{x\sigma},\nn
\vec{\hat{S}}_x&\equiv&\frac{1}{2}C^\dagger_x \vec{\sigma}C_x\ \ 
(\vec{\sigma}:{\rm Pauli\ matrices}),
\label{tJH}
\ee
where $C_{x\sigma}$ is the electron operator at the site $x$ 
satisfying the fermionic anticommutation relations.  
 $\sigma\ [\ =1(\uparrow), 2(\downarrow)]$ is the spin index
 and $\bar{1}\equiv 2, \bar{2}
 \equiv 1$ denote the opposite spin.
$\mu (=1,2,3)$ is the 3D direction index and also denotes
the unit vector.
The first $t$ term describes nearest-neighbor(NN) hopping of an 
electron without changing spin directions, i.e., ferromagnetic(FM)
hopping.
The second $J$ term describes the nearest-neighbor(NN) AF spin coupling 
of electrons. 
The doubly occupied states ($C^\dag_{x\uparrow}
C^\dag_{x\downarrow}|0\ra$) 
are excluded from the physical states due to the strong Coulomb
repulsion energy. 
The operator $\tilde{C}_{x\sigma}$ respects this point.
We adopt the {\em slave-fermion representation}
of the electron operator $C_{x\sigma}$ as a composite form,
\be
C_{x\sigma} = \psi^\dagger_x a_{x\sigma},
\ee 
where
$\psi_x$ represents annihilation operator of the 
 fermionic holon carrying the charge $e$ and no spin and
$a_{x\sigma}$ represents annihilation operator of the bosonic 
spinon carrying $s=1/2$ spin and no charge.
Physical states satisfy the following constraint, 
\begin{equation}
(\sum_\sigma 
a_{x\sigma}^\dagger a_{x\sigma} +\psi_x^\dag \psi_x)|{\rm phys}\rangle
 = |{\rm phys}\rangle.
\label{const1}
\end{equation}
In the salve-fermion representation, the Hamiltonian (\ref{tJH})
is given as 
\begin{eqnarray}
H&=&-t\sum_{x,\pm\mu}\psi_x^\dagger a_{x\pm \mu}^\dagger a_x\psi_{x\pm \mu}
\nonumber \\
&&+{J \over 4}\sum_{x,\mu}(a^\dagger\vec{\sigma}a)_{x+\mu}\cdot
(a^\dagger\vec{\sigma}a)_x.
\label{tJH2}
\end{eqnarray}

We employ the path-integral expression
of the partition function $Z = {\rm Tr}\exp(-\beta H)$
of the $t$-$J$ model
 in the slave-fermion representation, and introduce the complex number 
$a_{x\sigma}(\tau)$  and the Grassmann number $\psi_x(\tau)$
at each site $x$ and imaginary time 
$\tau \in [0,\beta\equiv (k_{\rm B} T)^{-1}]$.
The constraint (\ref{const1}) is solved\cite{CP1} 
by introducing
CP$^1$ spinon variable $z_{x\sigma}(\tau)$,
i.e., two complex numbers $z_{x1}, z_{x2}$ for each site $x$
satisfying 
\be
\sum_\sigma \bar{z}_{x\sigma} z_{x\sigma} = 1,
\label{z}
\ee
and writing 
\be
a_{x\sigma} = (1-\bar{\psi}_x\psi_x)^{1/2}z_{x\sigma}.
\label{a}
\ee
It is easily verified that the constraint (\ref{const1})
is satisfied by Eqs.(\ref{a}) and (\ref{z}).
Then, the partition function 
in the path-integral representation is given by
an integral over the
CP$^1$ variables $z_{x\sigma}(\tau)$ and Grassmann numbers 
$\psi_x(\tau)$.

The bosonic $t$-$J$ model in the slave-particle representation
is then defined at this stage\cite{hcb} 
by replacing $\psi_x$ by a U(1) boson field $\phi_x$ (Higgs field) as
\be
\psi_x \rightarrow \sqrt{\mathstrut \delta}\ \phi_x,\
\phi_x = \exp(i\varphi_x),
\label{replace}
\ee   
where $\delta(=\la \psi_x^\dag \psi_x\ra)$ 
is the hole concentration per site (doping parameter),
i.e., $\la \sum_\sigma C^\dag_{x\sigma} C_{x\sigma}\ra
= 1-\delta$.
Here we should mention that we have assumed uniform 
distribution of holes and set the amplitude in front of
 $\phi_x$ a constant ($\sqrt{\mathstrut \delta}$).
Validity of this assumption was partly supported by the 
numerical study of the closely related model with an
{\em isotropic} AF coupling\cite{uniform,btj}.
Actually, it is reported  there that
the ground state of the bosonic $t$-$J$ model (with $J_x=J_y=J_z$)
is  spatially uniform (without phase separation)
for $J/t \le 1.5$. 
We shall discuss on this point  further in Sec.6. 

We shall consider the system at finite and relatively high
$T$'s, such that
{\it the $\tau-$dependence of the variables $z_{x\sigma}, 
\phi_x$ are negligible} (i.e., only their zero modes survive).
Then the kinetic terms of $z_{x\sigma}, \phi_{x}$ including
 $\bar{z}_x\partial z_x/\partial \tau, 
 \bar{\phi}_x\partial \phi_x/\partial \tau$ disappear, and 
 the $T$-dependence may appear only 
 as an overall factor
$\beta$, which may be absorbed into the coefficients of the action 
and one may still deal with the 3D model
instead of the 4D model.

To obtain a description in terms of smooth spinon variables,
we change the CP$^1$ variables $z_{x\sigma}$ at {\em odd sites}
(the sites at which $x_1+x_2+x_3$ is odd) 
to the {\it time-reversed} CP$^1$ variable 
$\tilde{z}_{x\sigma}$ as
\be
z^{\rm old}_{x\sigma} &\rightarrow& z^{\rm new}_{x\sigma}=
\tilde{z}^{\rm old}_{x\sigma}\equiv
\sum_{\sigma'}\epsilon_{\sigma \sigma'}
\bar{z}^{\rm old}_{x\sigma'}\nn
&& {\rm for}\ \epsilon_x\equiv (-)^{x_1+x_2+x_3}=-1,
\label{newz}
\ee
where $\epsilon_{\sigma \sigma'}$ is the antisymmetric tensor
$\epsilon_{12}=-\epsilon_{21}=1, \epsilon_{11}=\epsilon_{22}=0$.
(Thus $\tilde{z}_{x1}=\bar{z}_{x2},\tilde{z}_{x2}=-\bar{z}_{x1}$.)
We stress that {\em this is merely a change of variables} 
in the path integral. (The AF spin configuration
in the original variables becomes a FM spin 
configuration in the new variables.) 

In this way, the partition function $Z$ of the 3D model at finite $T$'s 
is given by the path integral,
\be
Z=\int \prod_{x(x_1,x_2,x_3)}\big[dz_xd\phi_x
\prod_\mu dU_{x\mu}\big]\exp A,
\label{Z}
\ee
where the action $A$ on the 3D lattice is given by
\be
A&=&A_{\rm s}+A_{\rm h},\nn
A_{\rm s}&=&\frac{c_1}{2}\sum_{x,\mu}
\Big(\bar{z}_{x+\mu}U_{x\mu}z_{x} + \mbox{c.c.}\Big), \nn
A_{\rm h}&=&{c_3 \over 2}\left[
\sum_{x\in {\rm even},\mu}\tilde{z}_{x+\mu}\bar{z}_x 
\bar{\phi}_{x+\mu}\phi_x \right. \nn
&&\left.+\sum_{x\in {\rm odd},\mu}\tilde{z}_x\bar{z}_{x+\mu}\phi_{x+\mu}
\bar{\phi}_x+{\rm c.c.}\right],
\label{action}
\ee
where $\bar{z}'z\equiv\sum_{\sigma}\bar{z}'_\sigma z_\sigma$, 
etc.\cite{irrelevant}.
We have introduced the  U(1) gauge
field $U_{x\mu} \equiv \exp(i\theta_{x\mu})$ on the link $(x,x+\mu)\
(\mu=1,2,3)$
as an auxiliary field to make the action in a simpler form
and the U(1) gauge invariance manifest.
It corresponds to $U_{x\mu} 
\leftrightarrow \bar{z}_x z_{x+\mu}/|\bar{z}_x z_{x+\mu}|$.
In fact, one may integrate over $U_{x\mu}$ in  Eq.(\ref{Z}) to obtain
$\sum_{x\mu}\log I_0(c_1^2|\bar{z}_{x+\mu}z_{x}|)$ in the action
($I_0$ is the modified Bessel function),
which should be compared with the original expression
$\sum_{x\mu}|\bar{z}_{x+\mu}z_{x}|^2$.
Both actions have similar behavior and it is verified 
for $c_3=0$ that they  give rise to 
second-order transitions at similar values of $c_1$.

The action $A$ is invariant 
under a local ($x$-dependent) U(1) gauge transformation,
\be
z_{x\sigma} &\rightarrow& e^{i\lambda_x}z_{x\sigma}, \nn 
U_{x\mu}&\rightarrow& e^{i\lambda_{x+\mu}}U_{x\mu}e^{-i\lambda_x},\nn 
\phi_x &\rightarrow& e^{i\epsilon_x\lambda_x} \phi_x.
\ee
The gauge-invariant {\it bosonic} electron variable
$B_{x\sigma}$ is expressed as a composite as 
\be 
B_{x\sigma} =
\sqrt{\mathstrut \delta}\
\phi^\dag_x \times 
\left\{
\begin{array}{l}
z_{x\sigma}\ {\rm for}\ 
\epsilon_x=1,\\
\tilde{z}_{x\sigma}\ {\rm for}\ 
\epsilon_x=-1.
\end{array}
\right.
\ee
From Eqs.(\ref{tJH2}), (\ref{action}), and
the asymptotic form of $\log I_0(x)$,  
the parameters $c_1$ and $c_3$ are related with those in the
original $t$-$J$ model as 
\be
c_1&\sim& 
\left\{
\begin{array}{ll}
J\beta& {\rm for}\ c_1 >> 1,\\
(2J\beta)^{1/2}& {\rm for}\ c_1 << 1,
\end{array}
\right.
\nn 
c_3&\sim& t\delta\beta.
\label{c1c3}
\ee

In terms of holons and spinons, one of our motivations 
is put more explicitly.
In the slave-fermion representation of the fermionic $t$-$J$ model, 
the holon-pair field $\Phi_{x\mu} \equiv \psi_x\psi_{x+\mu}$ 
defined on the link 
behaves as a boson and a SC state appears as a result of 
its BEC.
Hoppings of $\Phi_{x\mu}$ are to destroy the magnetic order because of 
their couplings to spinons as
\be
A_{\rm SC} \sim \sum_{x,\mu,\nu}(\bar{z}_{x+\nu}\tilde{z}_x)\Phi_{x\mu}
(\bar{\tilde{z}}_{x+\mu}z_{x+\mu+\nu})\bar{\Phi}_{x+\nu,\mu},
\ee
which is similar to $A_{\rm h}$ in Eq.(\ref{action}).
The difference is that  $\phi_x$ is the site variable, while
 $\Phi_{x\mu}$ is the link variable.
Therefore, to study the CP$^1$ model coupled with $\phi_x$
may give us important insight to the dynamics of CP$^1$ spinons and 
holon  pairs on links,
in particular the interplay of the AF order and SC.

At the half filling $(\delta =0)$, only the spinon part 
$A_{\rm s}$ survives,
which describes the CP$^1$ model. The parameter
$c_1 \sim  J/(k_{\rm B}T)=J\beta$ controls fluctuations of $z_x$ and
$U_{x\mu}$.
The pure CP$^1$ model $A_{\rm s}$ exhibits a phase transition at 
$c_1=c_{1c}\sim 2.8$\cite{TIM1}.
In the low-$T$ phase($c_1 > c_{1c}$), the O(3) spin variable,
\be  
\vec{S}_x\equiv\bar{z}_x\vec{\sigma}z_x, \;\; \vec{S}_x\cdot \vec{S}_x=1,
\ee
made out of $z_{x\sigma}$
has a long-range order, $\lim_{|x|\rightarrow \infty}
\la \vec{S}_x \vec{S}_0\ra =m^2 \neq 0$,
i.e., the N\'eel order in the original model.
(Note that the replacement (\ref{newz})
leads to $\vec{S}^{\rm new}_x =\epsilon_x \vec{S}^{\rm old}_x$,
so an AF configuration of $\vec{S}^{\rm old}$
corresponds to a FM configuration of
$\vec{S}^{\rm new}_x$. Throughout the paper we use
the terms AF or FM configurations referring to the {\em original
spins} $\vec{S}^{\rm old}_{x}$.) 
The low-energy excitations are gapless spin waves, and the gauge
dynamics is in the Higgs phase. 
The high-$T$ phase($c_1 < c_{1c}$) is the  paramagnetic phase, where
the gauge dynamics is in the confinement phase and 
the lowest-energy excitations are spin-triplet bound states 
$\vec{S}_x$ of the spinon pairs.

\begin{figure}[b]
\vspace{0.7cm}
\begin{center}
\includegraphics[width=7.5cm]{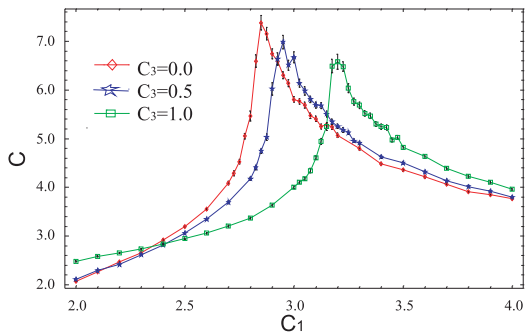}
\end{center}
\caption{
Specific heat for $L=24$ vs  $c_1$ for fixed $c_3$.
It exhibits a peak of AF transition 
at $c_1=2.8(2.95, 3.25)$ for $c_3=0(0.5, 1.0)$.}
\label{Fig:C_c1}
\end{figure}

Let us next consider the role of  the holon part $A_{\rm h}$.
$A_{\rm h}$ comes from the hopping $t$ term in (\ref{tJH}) 
and the parameter $c_3$ is expressed as in Eq.(\ref{c1c3}.
The spinon-pair amplitude of $z_x$ and $z_{x+\mu}$ in
$A_{\rm s}$ like
$\tilde{z}_{x+\mu}\bar{z}_x$ measures the NN
FM order of the original O(3) Heisenberg spins
due to the relation,
\be
|\bar{z}_{x+\mu}\cdot \tilde{z}_z|^2=
\frac{1}{2}(1-\vec{S}_x\cdot\vec{S}_{x+\mu} ).
\ee
Thus $A_{\rm h}$ favors {\it both
a FM coherent hopping amplitude, 
$\langle \bar{z}_{x+\mu}\tilde{z}_x \rangle$
and a BEC of $\phi_x$}. 
In other words, a BEC of $\phi_x$ requires a short-range FM
spin ordering.\\

\section{Numerical results I 
(3D model at finite $T$'s)}
\setcounter{equation}{0} 

In this section we report the results
of Monte Carlo  simulations that we performed
for the system given by Eqs.(\ref{Z}) and (\ref{action}).
We considered the cubic lattice with the periodic boundary
condition with the system size (total number of sites) 
$V=L^3$ up to $L=30$,
and used the standard Metropolis algorithm.
The typical statistics was $10^5$ MC steps per sample,
and the averages and errors were estimated over ten samples.
The typical acceptance ratio was about $50$\%.
In the action we added a very small but finite
($\sim 10^{-7}$) external magnetic field.

Let us start with the region of relatively high $T$'s.
In Fig.\ref{Fig:C_c1} we show the specific heat per site $C$,
\be
C=\frac{1}{V}\langle (A-\langle A \rangle)^2\rangle,
\label{c3d}
\ee
as a function of $c_1$ for several values of $c_3$.
$C$ exhibits a peak that
shifts to larger $c_1$  as $c_3$ is increased.
This transition is nothing but the AF N\'eel phase transition
observed previously for the $c_3=0$ case\cite{TIM1}.
Fig.\ref{Fig:C_c1} and the relations (\ref{c1c3}) show 
that the N\'eel temperature is lowered 
by doped holes as we expected.

\begin{figure}[t]
\begin{center}
\includegraphics[width=7.5cm]{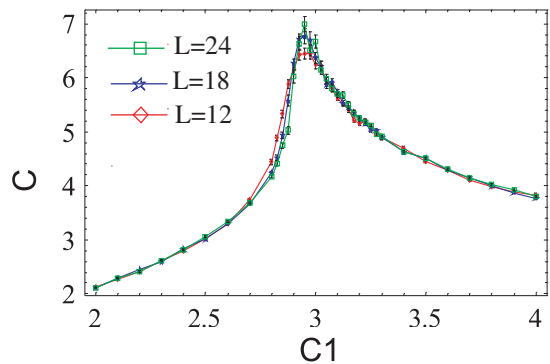}
\end{center}
\caption{
Size-dependence of the specific heat for $c_3=0.5$.}
\label{Fig:sizedepc}
\end{figure}

\begin{figure}[t]
\begin{center}
\includegraphics[width=8cm]{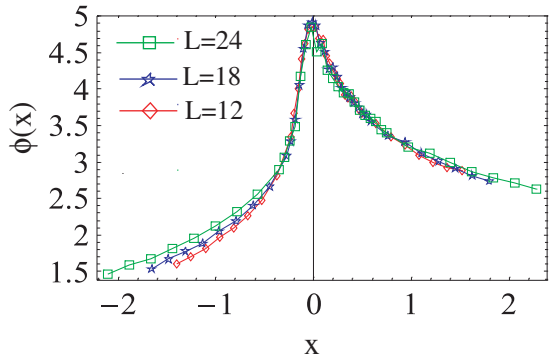}
\end{center}
\caption{
Scaling function $\phi(x)$ of (\ref{scaling}) 
determined from the data of Fig.\ref{Fig:sizedepc}.
}
\label{Fig:scaling}
\end{figure}

In Fig.\ref{Fig:sizedepc} we present the system-size dependence 
of $C$ for $c_3=0.5$ in which
the specific heat $C$ develops 
moderately but systematically as we increase $L$.
We fitted this $C$ by using the finite-size scaling(FSS) of the form,
\be
C(c_1,L)=L^{\sigma/\nu}\phi(L^{1/\nu}\epsilon),
\label{scaling}
\ee
where $\epsilon=(c_1-c_{1c})/c_{1c}$ with the critical coupling
$c_{1c}$ at infinite system size, and $\sigma, \; \nu$ are
critical exponents.
In Fig.\ref{Fig:scaling} we show the determined 
scaling function $\phi(x)$, from which we 
estimated the critical exponent 
of correlation length as $\nu=1.7$.

\begin{figure}[b]
\begin{center}
\includegraphics[width=7cm]{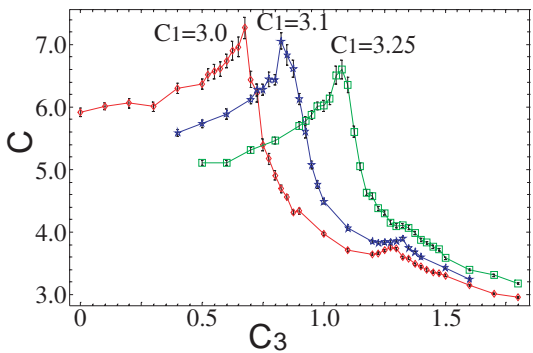}
\end{center}
\caption{
Specific heat for $L=24$ vs $c_3$ for fixed $c_1$.
There are two peaks, which indicate the AF 
transition(in the smaller $c_3$-region) and the BEC transition
(in the larger $c_3$ region).
}
\label{Fig:C_c3_high_t}
\end{figure}


In Fig.\ref{Fig:C_c3_high_t} we present $C$ as a function of $c_3$ 
for fixed $c_1$.
There exist two peaks in $C$.
To study the origin of these peaks, we define
the ``specific heat" $C_{\rm s}$ and $C_{\rm h}$
for each term   $A_{\rm s}$ and $A_{\rm h}$  
of the action (\ref{action}) separately as

\begin{figure}[t]
\begin{center}
\vspace{-2cm}
\includegraphics[width=7cm]{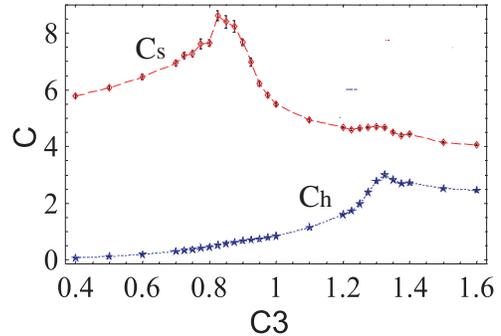}
\end{center}
\caption{
Specific heat $C_{\rm s}$ and $C_{\rm h}$
of (\ref{csh3d})
at $c_1=3.1$ and $L=24$.
The location of peak of $C_{\rm s(h)}$ 
almost coincides the location of peak of $C$ in
Fig.\ref{Fig:C_c3_high_t} at smaller(larger) $c_3$.
}
\label{Fig:eachc}
\end{figure}


\be
C_{\rm s,h}=\frac{1}{V}\langle (A_{\rm s,h}-
\langle A_{\rm s,h} \rangle)^2\rangle.
\label{csh3d}
\ee
In Fig.\ref{Fig:eachc} we present $C_{\rm s}$ and $C_{\rm h}$
for $c_1=3.1$, in which $C_{\rm s}$ has a peak at $c_3 \simeq 0.8$,
the smaller-$c_3$ peak position of $C$ while 
$C_{\rm h}$ has a peak at $c_3 \simeq 1.3$,
the larger-$c_3$  position of $C$. 
From this result, we identify the peak at smaller $c_3$
expresses the AF transition in Fig.\ref{Fig:C_c1},
which is  generated by $A_{\rm s}$,
and 
the peak at larger $c_3$ expresses the BEC transition,
which is  driven by $A_{\rm h}$.
Because each peak develops as $L$ is increased, 
they are both second-order phase transitions.
The critical exponent of $C_{\rm s}$ of Fig.\ref{Fig:eachc} 
is estimated as $\nu=0.70$.

\begin{figure}[b]
\begin{center}
\includegraphics[width=7.5cm]{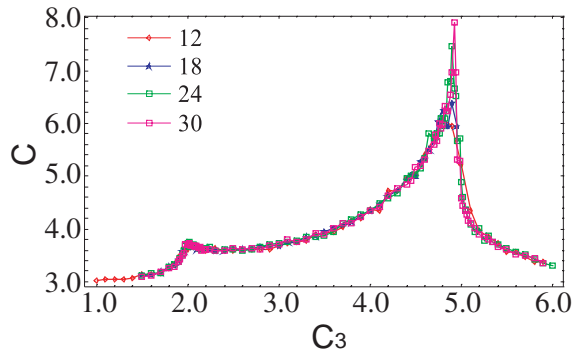}
\end{center}
\caption{
$C$ vs $c_3$ for $c_1=6.0$ with $L=12, 18, 24, 30$.
There exist again two peaks of AF and BEC phase transitions.
However, their order are reversed (BEC peak in smaller $c_3$
and AF peak at larger $c_3$).
Both peaks develop as $L$ is increased,
supporting that they are of second order.
In the region between the two peaks, the AF order and
the BEC order coexist.  
}
\label{Fig:C_c3_low_t}
\end{figure}

Let us turn to the low-$T$ (large-$c_1$) region and see what happens
to the AF and BEC phase transitions.
In Fig.\ref{Fig:C_c3_low_t}, we present $C$ for $c_1=6.0$.
Again there exist two peaks but the order of them is interchanged.
Both peaks develop as $L$ is increased, and
therefore both of them are still second-oder phase transitions.
As $c_3$ (or equivalently $\delta$) is increased, 
the transition into the BEC phase takes place
first and then the AF phase transition follows.
This means that there exists a phase in which
both the AF and BEC long-range
orders coexist. 

To verify the above conclusion, we measured the spin-spin
correlation $G_{\rm AF}(r)$ of $\vec{S}_x$
and the correlation $G_{\rm BEC}(r)$ of the gauge-invariant
``composite electron" pair variable $V_{x\mu}$
in each phase,
\begin{eqnarray}
G_{\rm AF}(r)&=&{1 \over 3V}\sum_{x,\mu}
\langle \vec{S}_{x+r\mu}\cdot \vec{S}_x \rangle,   \label{G_S} 
\ee
\be
G_{\rm BEC}(r)&=&{1 \over 12V}\sum_{x,\mu\neq \nu}
       \langle \bar{V}_{x+r\nu,\mu}V_{x\mu} \rangle+\mbox{c.c.},
       \label{G_SC}\\
V_{x\mu}&\equiv&
\phi_{x+\mu}\phi_x\times\left\{
     \begin{array}{ll}
      z_{x+\mu}\bar{z}_x, \;\;
      & \epsilon_x= 1, \\
      \bar{z}_{x+\mu}z_x, & \epsilon_x =-1 .
     \end{array}
\right.\nonumber     
\end{eqnarray}

\begin{figure}[t]
\begin{center}
\hspace{-2cm}
\includegraphics[width=9cm]{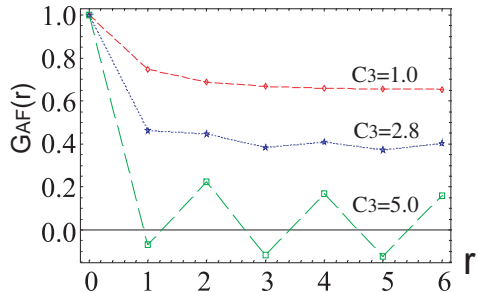}
\end{center}
\caption{\label{Fig:spin_c1=6}
Spin-spin correlation function $G_{\rm AF}(r)$ of
(\ref{G_S}) in the three phases along $c_1=6.0$.
For small $c_3$, $G_{\rm AF}(r)$
 has a nonvanishing smooth component, i.e.,
 the genuine AF staggered magnetization.
(Remember the direction of the spin on the odd sites has
been reversed.)
For intermediate $c_3$, there appears an oscillatory,
i.e., FM component in the AF background.
For larger $c_3$, SC exists without AF long-range order, 
where the  FM long-range order dominates.
}\end{figure}

\begin{figure}[b]
\begin{center}
\includegraphics[width=7cm]{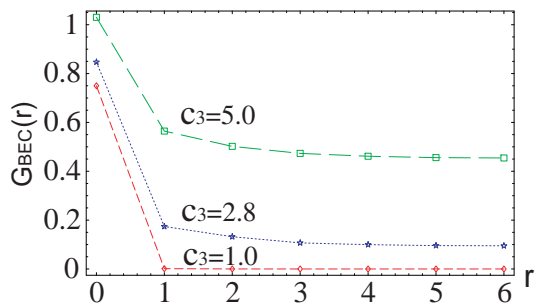}
\end{center}
\caption{\label{Fig:SC_c1=6}
Correlation function $G_{\rm BEC}(r)$ (\ref{G_SC})
of bosonic hole(vacant electron) pair $V_{x\mu}$ in the three phases
for $c_1=6.0$.
}\end{figure}

The results are shown in Figs.\ref{Fig:spin_c1=6} and
\ref{Fig:SC_c1=6}.
Fig.\ref{Fig:spin_c1=6} exhibits an interesting result 
in the coexisting phase of AF order and BEC  at intermediate 
$c_3(\propto \delta)$.
There the spin correlation has a FM 
component in the AF background. 
This shows that a short-range FM order is needed for 
the BEC
as some mean-field theoretical studies of the $t$-$J$ 
model\cite{mft} indicate.
As $c_3$ is increased further, the AF long-range order disappears
and the FM order appears instead as a result of the holon-hopping 
amplitude. Fig.\ref{Fig:SC_c1=6} shows that the BEC
certainly develops for larger $c_3$'s.

Let us comment here on the nature of BEC we have studied.
To study BEC we have used  $G_{\rm BEC}$, the 
  long-range order of which implies a 
BEC of {\it electron pairs}.
There may be another kind of BEC, a condensation 
of single bosonic electrons, which is measured by 
the electron-electron correlation function,
$\la \bar{B}_{x+r,\sigma}B_{x\sigma'}\ra$.
(Note that $\la \bar{\phi}_{x+r}\phi_{x}\ra$ is not suitable
because it is not gauge invariant and always vanishes.)
However, as we have seen in 
Figs.\ref{Fig:C_c3_high_t}-\ref{Fig:C_c3_low_t},
there is only one peak at most in the specific heat
apart from the peak representing the magnetic order,
so the two condensations, one  of single electrons and
the other of pairs of electrons
take place simultaneously.

\begin{figure}[t]
\begin{center}
\includegraphics[width=7cm]{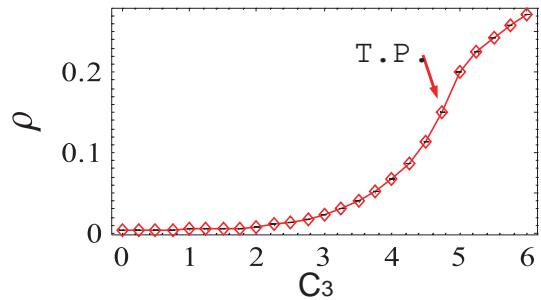}
\end{center}
\caption{\label{Fig:inst_c1=6}
Monopole density $\rho$ for $c_1=6.0$ and $L=24$.
Near the AF phase transition point at $c_3=4.8$
(with an arrow), 
the  curvature of $\rho$ changes its sign.
}\end{figure}


We  measured also the monopole density $\rho$\cite{TIM1}, 
which gives information on strongness of the fluctuation of the
gauge field $U_{x\mu}$.
In Fig.\ref{Fig:inst_c1=6} we present $\rho$ along $c_1=6.0$.
$\rho$ changes its behavior around 
the AF transition point at $c_3=4.8$,
indicating that the AF ordered phase
corresponds to the deconfinement phase of $U_{x\mu}$.
The low-energy excitations there are  gapless spin waves
described by the uncondensed component of $z_x$.
In the paramagnetic phase, on the other hand,
the confinement phase of $U_{x\mu}$ is realized and the 
low-energy excitations 
are the spin-triplet $\vec{S}_x$ with a gap.

\begin{figure}[t]
\vspace{-1.5cm}
\begin{center}
\includegraphics[width=6.2cm]{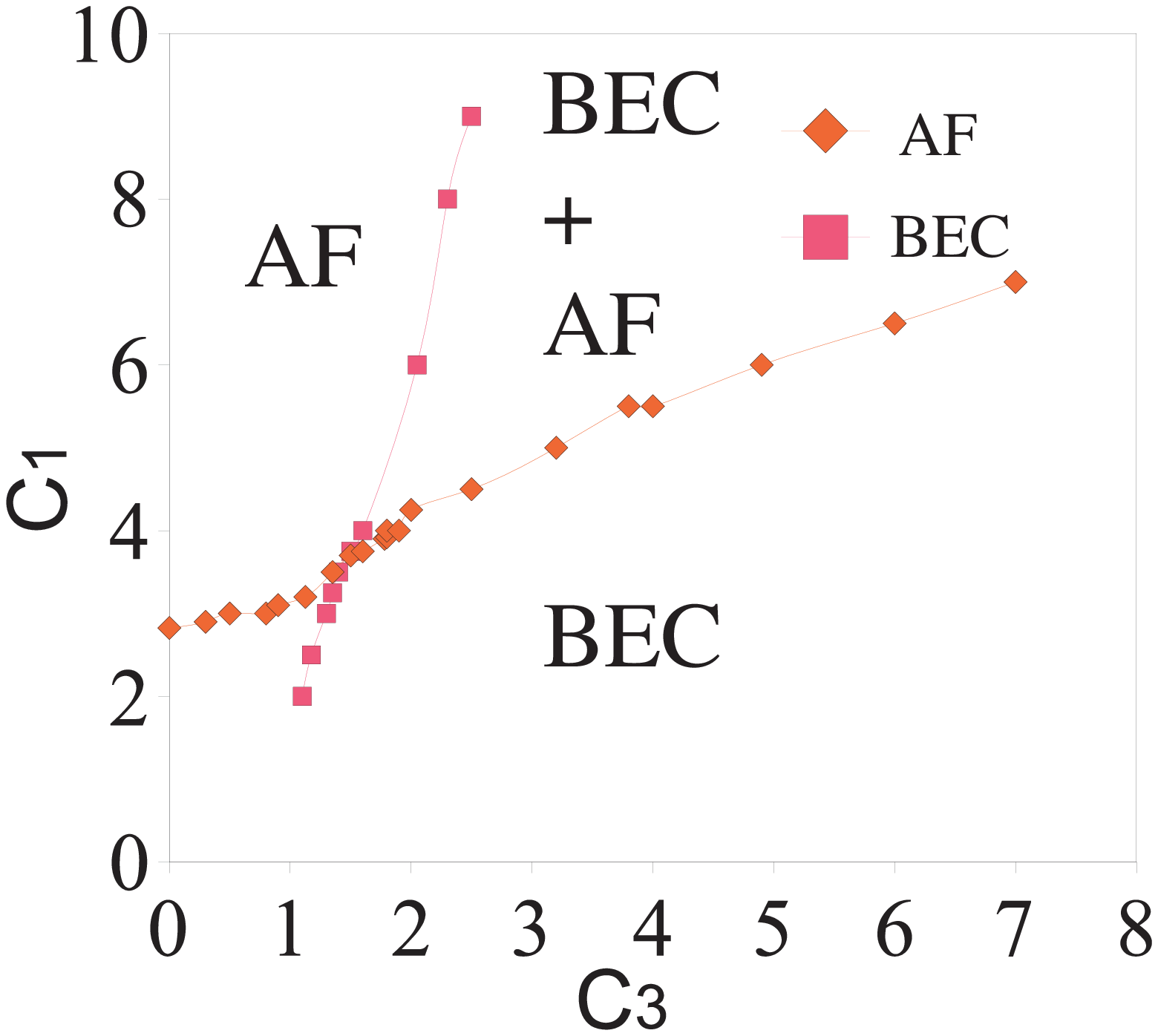}\\
(a)\\
\includegraphics[width=6.2cm]{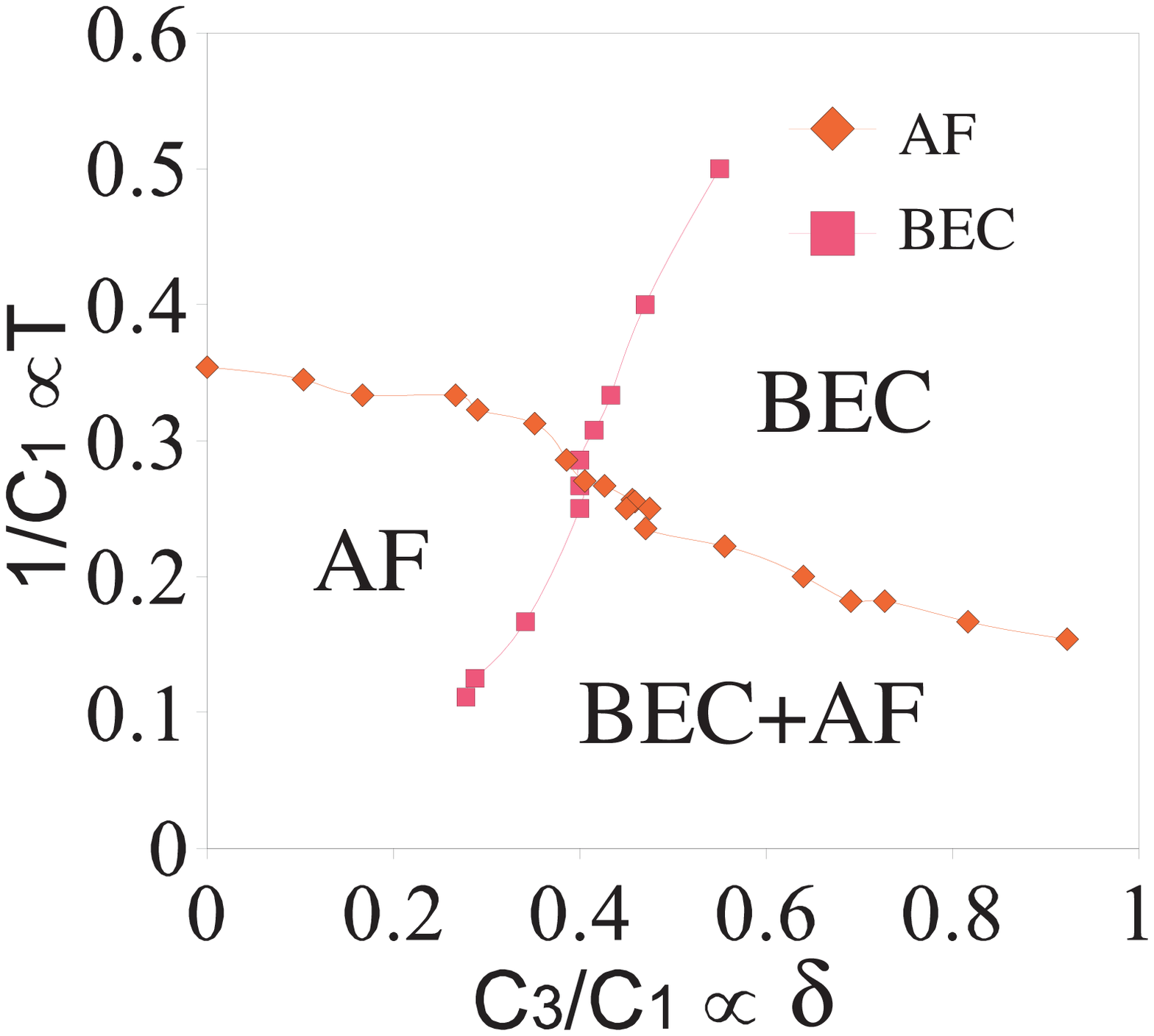}\\
(b)
\end{center}
\vspace{0.5cm}
\caption{\label{Fig:phase_T}
Phase diagram of the 3D model at finite $T$
in the $c_3-c_1$plane (a) and (b) in the
$c_3/c_1-1/c_1$ plane ($c_3/c_1 \propto 
t\delta/J$ and $1/c_1 \propto T$).
In the low-$T$ and intermediate $\delta$ region, there exists the 
 coexisting phase of BEC and AF order.
All the phase transition lines are of second order.
}\end{figure}

To summarize the results of the 3D system, we present   
in Fig.\ref{Fig:phase_T} the phase  diagram in the $c_3/c_1-1/c_1$,
i.e., $\delta-T$ plane. The N\'eel temperature 
decreases slowly as $\delta$ increases, while the BEC
critical temperature develops rather sharply. 
The two orders can coexist at low $T$'s in
intermediate region of $\delta$. 
Fig.\ref{Fig:phase_T} has a close resemblance to  
the phase diagram of Ref.\cite{btj}. 
However, in Ref.\cite{btj}, the anisotropic spin coupling
is considered and the phase separation  occurs as a result.
The coexistence phase of the AF and BEC orders
obtained in Ref.\cite{btj} is nonuniform and accompanied with 
this phase separation.
In the present paper, we studied the system with 
the isotropic spin coupling, and 
the phase with both the AF and BEC orders is realized under the
uniform distribution of holes\cite{su2}.
In this sense, the model in the present paper is close to
the fermionic $t$-$J$ model with parameters $t>J$\cite{sorella}.
We expect that the fermionic $t$-$J$ model has a similar phase diagram 
to Fig.\ref{Fig:phase_T}\cite{btJ2}.
In Sec.6, we explain further the implication of the results obtained in 
the present paper to the phase structure of the fermionic $t$-$J$ model.

\section{Bosonic $t$-$J$ model in the CP$^1$-Higgs representation:
2D model at $T=0$}
\setcounter{equation}{0} 

As mentioned in Sect.1, the cold atoms can be put on a 
2D optical lattice. At $T=0$ one may expect
a BEC and/or magnetic ordering at certain conditions for
density of atoms per well,
interaction between atoms, etc.
We expect that the bosonic $t$-$J$ model in the present section 
describes
dynamics of bosonic atoms with (pseudo-)spin degrees of freedom 
and (strong) repulsive interactions between them.
For the physics of high-$T_{\rm c}$ cuprates, it is also interesting
to study the two-dimensional (2D) bosonic $t$-$J$ model at $T=0$.
The main concern there is the  {\em quantum phase transitions} 
(QPT) of magnetism, MIT, and SC.
Study of the QPT in the doped CP$^1$ model is not only important for 
verifying the phase diagram at finite $T$ obtained in 
the above, but also interesting for physics of cold atom systems
in optical lattices.

In this section, we shall briefly review 
the derivation of the effective
field-theory model for the 2D fermionic $t$-$J$ model 
at $T=0$\cite{CP1}. Then, in this effective model,
by replacing the fermionic holon variables
by Higgs boson variables, we
obtain the bosonic effective model of the 
2D bosonic $t$-$J$ model at $T=0$.
The main difference from
the previous sections (the 3D model at finite $T$'s) 
is that the action contains
the kinetic terms of spinons and holons due to their
nontrivial $\tau$-dependence, and one should
consider the 2+1 dimensional model.
They are both 3D models but
the couplings along the third-direction ($x_3$ or $\tau$)
are  different in the two models.

Let us start with the path-integral representation
of  the (quantum) partition function $Z_{\rm F}^{\rm 2D}$ 
of the 2D fermionic $t$-$J$ model in the 
slave-fermion representation
with the CP$^1$ variables\cite{CP1},
\be
Z_{\rm F}^{\rm 2D}=\int \prod_{r,\tau}[
d\psi_{r}(\tau) 
dz_{r}(\tau)]
\exp\Big( A^{\rm 2D}_{\rm F}(\tau)\Big),
\label{ZT=0}
\ee
where $r=(x_1, x_2)$ denotes the site of the 2D lattice 
and $\tau \in (0, \infty)$ is the imaginary time.
The action $A^{\rm 2D}_{\rm F}(\tau)$ 
is given by\cite{irrelevant}
\begin{eqnarray}
A^{\rm 2D}_{\rm F}(\tau)&=&\int_0^{\infty}d\tau\sum_{r}
\Big[-\big( \bar{\psi}_{r}D_{r\tau}\psi_{r}
+\bar{z}_{r}\partial_\tau z_{r} \big)  \nonumber  \\
&-&t\sum_{i=1,2}
\big(\bar{\psi}_r\bar{z}_{r+i}z_r\psi_{r+i}
+\bar{\psi}_{r}\bar{z}_{r-i}z_{r}\psi_{r-i}
\big)
\nn
&-&{J\over 4}\sum_{i}\bar{z}_{r}
\vec{\sigma}z_{r}\cdot
\bar{z}_{r+i}\vec{\sigma}z_{r+i}\Big], \label{AT=0}\\
D_{r\tau}&\equiv&\partial_\tau+iA_{r\tau},\ 
A_{r\tau}\equiv i \bar{z}_r\partial_\tau z_r,
\end{eqnarray}
where $i=1,2$ denotes the spatial direction index and 
also the unit vector.

Each kinetic term, $\bar{\psi}_rD_{r\tau} \psi_r$,
$\bar{z}_r\partial_{\tau} z_r$, in $A^{\rm 2D}_{\rm F}$
is purely imaginary, so the 
straightforward MC simulations cannot be applied.
However, concerning to $\bar{z}_r\partial_{\tau} z_r$,
by integrating out the half of the CP$^1$ variables
(e.g., $z_r$'s on all the 2D odd sites, i.e.,
 $\epsilon^{\rm 2D}_x\equiv(-)^{x_1+x_2}=-1$) 
one may obtain a purely real action as a result.
That real action describes the low-energy spin excitations 
in a natural and straightforward manner.

Explicitly, we assume  a short-range AF order,
\begin{equation}
(\bar{z}\vec{\sigma}z)_r\sim -(\bar{z}\vec{\sigma}z)_{r\pm i}.
\label{SRAF}
\end{equation}
Then we parameterize each CP$^1$ variable $z_{\rm o}$ at 
the  2D odd site o 
by referring to one of its NN partner $z_{\rm e}$ at the
even site e at equal time as
\begin{equation}
z_{\rm o}=p_{\rm oe}
z_{\rm e}+(1-|p_{\rm oe}|^2)^{1/2}U_{\rm oe}\tilde{z}_{\rm e},
\label{para}
\end{equation}
where $p_{\rm oe}$ is a complex number sitting
on the link (oe) and $U_{\rm oe}$ is a U(1) phase 
factor that
makes the parameterization (\ref{para}) to be consistent with 
the local gauge symmetry\cite{voe}.
The assumption (\ref{SRAF}) implies $|p_{\rm oe}|^2 \ll 1$,
which is justified by the $J$-term in the 
action (\ref{AT=0}) for the lightly doped case ($\delta \ll1$)\cite{sraf}.
Integration over $z_{\rm o}$ is reduced
to the integral over $p_{\rm oe}$, 
which may be approximated  as the Gaussian integration  link by link,
\be
&&\int_{-\infty}^{\infty} d\bar{p}dp
\exp(-\bar{p}Mp+\bar{p}k+\bar{\ell}p)\nn
&&\ \ \ \propto
{\rm det}M^{-1}\exp(\bar{\ell}M^{-1}k).
\ee

The CP$^1$ part  of the resulting
 action  is not symmetric concerning to spatial directions
because the choice of a definite even-site partner
breaks it. However, by considering the smooth 
configurations of spinons $z(r,\tau)$,
one recovers the symmetric action in the form of
continuum space\cite{acp1},
\be
A_{{\rm CP}^1}&=&-{1 \over 2a^2}\int_0^{\infty}d\tau\int d^2r\nn
&\times&
\Big(Ja^2\sum_{i=1,2} \overline{D_i z}D_i z
+{1 \over 2J}\overline{D_\tau z}D_\tau z \Big),\nn
D_{\mu} &\equiv& \partial_\mu+iA_{\mu},\ 
A_{\mu}\equiv i\bar{z}\partial_\mu z,
\label{ACP1}
\ee
where $a$ is the lattice spacing (hereafter we often set $a=1$).
The relativistic couplings of (\ref{ACP1})
give rise to spin wave excitations in the background of
AF order with the dispersion $E(k) \propto k$
instead of $E(k) \propto k^2$ in the FM spin model.

The integration over the odd-site spins affects the 
holon and spinon hopping term.
The most important point is that the U(1) factor $U_{\rm oe}$
always appears in combination with $\psi_{\rm o}$ as
$\bar{U}_{\rm oe}\psi_{\rm o}$, and therefore we redefine 
$\bar{U}_{\rm oe}\psi_{\rm o}\rightarrow \psi_{\rm o}$.
Then new $\psi_{\rm o}$ transforms under a gauge transformation as
$\psi_{\rm o}\rightarrow e^{-i\theta_{\rm e}}\psi_{\rm o}$. 
This property is consistent with the fact that spinon hopping
amplitude, e.g., from even site (e') to odd site (o), contains factor
$(\tilde{z}_{\rm e}\cdot \bar{z}_{{\rm e}'})$ in the resultant
effective action like
\begin{equation}
(\tilde{z}_{\rm e}\cdot \bar{z}_{{\rm e}'})
\bar{\psi}_{\rm o}\psi_{{\rm e}'}+\mbox{c.c.}
\label{hop2}
\end{equation}

Finally, the holon kinetic term is now expressed  by using 
$\psi_{\rm o}^{\rm new} = \bar{U}_{\rm oe}\psi_{\rm o}^{\rm old}$ as\cite{eta}
\be
-\int d\tau\left[
\sum_{\rm e} \bar{\psi}_{\rm e}(\partial_\tau+iA_{{\rm e}\tau}) 
\psi_{\rm e}
+\sum_{\rm o} \bar{\psi}_{\rm o}(\partial_\tau-iA_{{\rm e}\tau}) 
\psi_{\rm o}
\right].\nn
\label{holonkinetic}
\ee

To proceed  we note  that the obtained total action 
is not symmetric w.r.t. the spatial directions,
but a symmetric one can be available
by reintroducing CP$^1$ variables at the 2D odd sites
according to the invariance under a local U(1) gauge transformation.
Furthermore,
by choosing the lattice spacing in the $x_3$ direction
suitably, the symmetry  in all the three
directions is recovered.
Then the spin part $A^{\rm 2D}_{\rm s}$ of the action
becomes nothing but the 3D action $A_{\rm s}$ of (\ref{action}).
Both of them have the same continuum limit (\ref{ACP1}) 
for the spin part, so they are 
to be categorized to the same universality class.

Then the bosonic $t$-$J$ model on the lattice is obtained by the replacement
$\psi_x \rightarrow \sqrt{\mathstrut \delta}
\phi_x, \ \phi_x \equiv  \exp(i\varphi_x)$ as in Eq.(\ref{replace}).
The partition function $Z^{\rm 2D}$ on the 2+1-dimensional spacetime
lattice is then given by
\be
Z^{\rm 2D}&=&\int\prod_x[dz_xd\phi_x\prod_\mu dU_{x\mu}]\exp(A^{\rm 2D}),\nn 
A^{\rm 2D}&=&A^{\rm 2D}_{\rm s}+A^{\rm 2D}_{\rm h},\ \ 
A^{\rm 2D}_{\rm s}=A_{\rm s}\ {\rm of\ Eq.}(\ref{action}),
\nn
A^{\rm 2D}_{\rm h}&=&{s_3 \over 2}\Big(
\sum_{x\in {\rm e}}
\bar{\phi}_{x+3}U_{x3}\phi_x 
+\sum_{x\in {\rm o}}
\bar{\phi}_{x+3}\bar{U}_{x3}\phi_x  \label{action2d}
\label{A2+1d}\\
&+&\sum_{x\in {\rm e},i}\tilde{z}_{x+i}\bar{z}_x 
\bar{\phi}_{x+i}\phi_x  
+\sum_{x\in {\rm o},i}\tilde{z}_x\bar{z}_{x+i} \phi_{x+i}
\bar{\phi}_x +{\rm c.c.}\Big).\nonumber
\ee
Here $x=(x_1,x_2,x_3)$, and $r=(x_1, x_2)$ 
stands for the coordinates of the 2D plane
and $x_3 (=0, 1,\cdots, \infty)$ 
is the discretized imaginary time ($\tau = x_3 a_3$). 
Even (e) and odd (o) sites are defined regarding to
 the {\em 2D plane} as 
$\epsilon^{\rm 2D}_x\equiv(-)^{x_1+x_2}=1$(even), -1(odd).

The coefficients $c_1, s_3$ are related to the parameters of 
the $t$-$J$ model as follows;\cite{c1s3}
\be 
c_1 &\sim& {\rm  constant\ independent\ of}\ J\ {\rm and}\ t,
\nn
s_3 &\sim& \frac{t}{J}\delta.
\label{s3}
\ee
$c_1$
measures the solidity of the
N\'eel state in the AF Heisenberg magnet at $T=0$.
$s_3$ is the hopping amplitude of electrons.
For the AF Heisenberg model with the standard NN exchange
coupling, $c_1>c_{1c}\simeq 2.8$, and the critical value $c_{1c}$
decreases (increases) if long-range and/or anisotropic
couplings that enhance (hinder) the AF order are added\cite{YAIM}.

In $A_{\rm h}$ we added the Hermitian conjugate 
of the  holon kinetic term,
$\bar{\phi}_{x+3}(\bar{U}_{x3}, U_{x3})\phi_x$, 
which were absent
in Eq.(\ref{holonkinetic}), to make the action 
real, explicitly.
This modification might sound crucial.
However, we expect it a modest replacement because the omitted imaginary 
part, $-i(\bar{\phi}_{x+3}U_{x3}\phi_x-$c.c.), etc, 
should have vanishing expectation value and 
behave mildly for the relevant configurations
for the original action without the complex conjugates.
Another support for this modification
is given by taking into account the fluctuation
of holon field\cite{rho}.

Let us summarize the difference between the 2D model at $T=0$
and the 3D model at $T>0$ of (\ref{action}).
To see it explicitly, it is convenient to 
use the redefinition, 
\be
\phi_x' \equiv \{
\begin{array}{l}
\bar{\phi}_x\ {\rm for}\ \epsilon^{\rm 2D}_x=-1\\
\phi_x\ {\rm for}\ \epsilon^{\rm 2D}_x=1 
\end{array}
({\rm 2D\ model}),
\ee
and the relation $\bar{z}\tilde{z}'= -\tilde{z}\bar{z}'$.
Then $A^{\rm 2D}_{\rm h}$ 
becomes
\be
A^{\rm 2D}_{\rm h}&=&{s_3 \over 2}\sum_{x}\Big(
\bar{\phi}'_{x+3}U_{x3}\phi_x' \nn
&+&\sum_{i}\epsilon^{\rm 2D}_x\tilde{z}_{x+i}\bar{z}_x 
\phi_{x+i}'\phi_x'+ {\rm c.c.}  
\Big).
\label{action2dv2}
\ee
In the similar notation, 
\be
\phi_x' \equiv \{
\begin{array}{l}
\bar{\phi}_x\ {\rm for}\ \epsilon_x=-1\\
\phi_x\ {\rm for}\ \epsilon_x=1
\end{array}
({\rm 3D\ model}),
\ee
the action $A_{\rm h}$
of the 3D model (\ref{action})
is given as
\be
A_{\rm h}&=&{c_3 \over 2}\sum_{x,\mu}\Big(
\epsilon_x\tilde{z}_{x+\mu}\bar{z}_x 
\phi_{x+\mu}'\phi_x'+ {\rm c.c.}  
\Big).
\label{action3dv2}
\ee
Thus, the hopping of holons $\phi_x$ along $\mu=3$ is 
uniform in the $\bar{\phi}\phi$-channel in the 2D model,
while it is alternative in the $\phi\phi$-channel 
in the 3D model, i.e., the latter has 
an extra factor $(-)^{x_3}$.
\vspace{1cm}

\section{numerical results II
 (2D model at $T=0$)}
\setcounter{equation}{0} 

\subsection{Phase diagram}

\begin{figure}[t]
\begin{center}
\includegraphics[width=7cm]{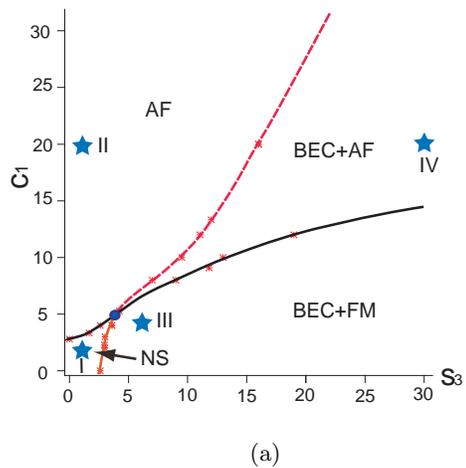}\\
\vspace{-1.3cm}
(a)\\
\hspace{-1.0cm}
\includegraphics[width=7cm]{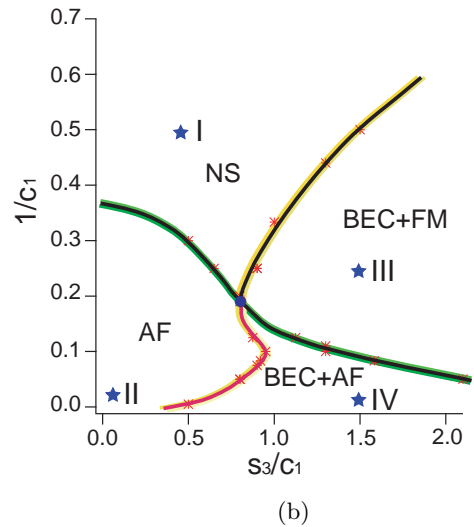}\\
(b)
\end{center}
\vspace{0.5cm}
\caption{\label{Fig:phase_T=0}
Phase diagram of the 2D $t$-$J$ model at $T=0$ in the 
$s_3-c_1$ plane (a) and (b) in the $s_3/c_1-1/c_1$ plane.
The $c_1$-term controls the stability of the AF N\'eel order,
whereas the $s_3$-term represents holon hopping.
In the large-$c_1$ and intermediate $s_3/c_1$, there exists the 
coexisting phase of AF order and BEC. At the merging point 
of two transition lines, the peak
in $C^{\rm 2D}$ disappears. 
NS stands for normal state and FM for ferromagnetic order.
The phase transition line separating AF and BEC+AF phases is of first order,
while other three transition lines are of second order. 
The four points (I-IV) marked by stars are selected
as typical points:
(I) $c_1=2, s_3=1;$\ (II) $c_1=20, s_3=1;$\ 
(III) $c_1=4, s_3=6;$\ (IV) $c_1=20, s_3=30.$\
}\end{figure}

We studied the quantum system $A^{\rm 2D}$ of (\ref{action2d})
by MC simulations\cite{mc3d}.
In this section we show the results of these calculations.
Let us first present the obtained phase diagram in Fig.\ref{Fig:phase_T=0}.
Its phase structure is globally 
similar to  Fig.\ref{Fig:phase_T} for  
the 3D model at finite $T$'s.
As $s_3$ is increased, the AF order is destroyed and
the BEC appears.
The appearance of a first-order transition line is 
new. Below we shall see the details of various quantities,
which support Fig.\ref{Fig:phase_T=0}.
We select the following four points in the phase diagram
to represent each of four phases;\\
\noindent
(I) $s_3=1, c_1=2$;\ Normal and paramagnetic state.\\
\noindent
(II) $s_3=1, c_1=20;$\ AF state without BEC. \\
\noindent 
(III) $s_3=6, c_1=4$;\ BEC state with a FM order\\
\noindent 
(IV) $s_3=30, c_1=20$;\ AF and BEC state.\\
\noindent 
As presented later, various quantities 
are measured on these points and compared with each other.

Let us comment here on the assumption 
of the short-range AF order (\ref{SRAF})
in deriving the effective model (\ref{A2+1d}). 
As the phase diagram of Fig.\ref{Fig:phase_T=0} shows, 
we find that there exists a FM order for small $c_1$ and large  $s_3$.
So the applicability of (\ref{SRAF})
in this parameter region might be questionable.
However, the existence of the FM order itself in Fig.\ref{Fig:phase_T=0} 
is consistent with  the results of the 3D finite-$T$ system
obtained in Secs.2,3 (as long as each phase at 
$T=0$  smoothly  continues to $T > 0 $).
Thus we expect that the global phase structure of
Fig.\ref{Fig:phase_T=0} is not affected by the assumption (\ref{SRAF}),
although the location and the nature of the phase boundaries 
need some modifications.
Furthermore, calculations of 
various physical quantities in that phase give us very important
insight in understanding the whole phase structure of the model.

\subsection{Specific heat and internal energy}

\begin{figure}[t]
\begin{center}
\hspace{-2.0cm}
\includegraphics[width=10.5cm]{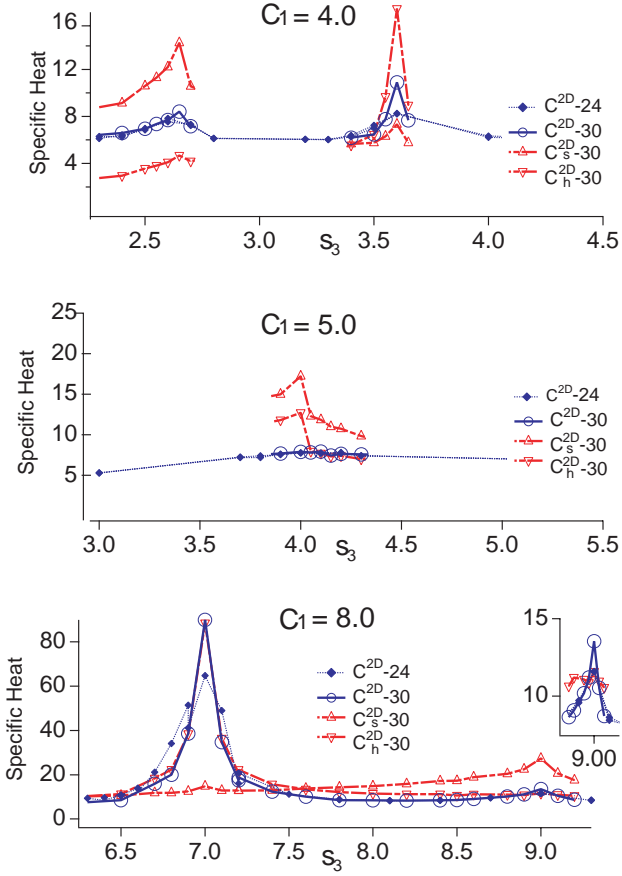}
\end{center}
\vspace{-0.5cm}
\caption{
``Specific heats", $C^{\rm 2D}, C^{\rm 2D}_{\rm s}, C^{\rm 2D}_{\rm h}$
vs  $s_3$ for $c_1=4.0, 5.0, 8.0$.
The notation $C_{\rm s}-24$ denotes $C^{\rm 2D}_{\rm s}$ 
for $L=24$, etc.
At $c_1=5.0$ the two peaks almost merge.}
\label{specificheats}
\end{figure}

\begin{figure}[t]
\begin{center}
\includegraphics[width=6.5cm]{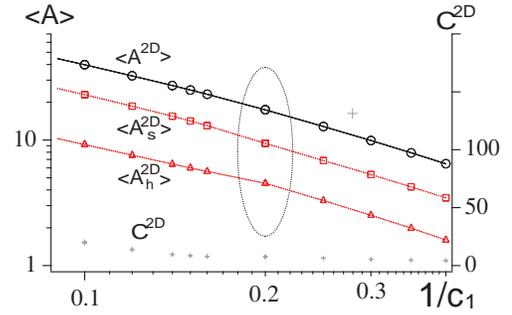}
\end{center}
\vspace{-0.5cm}
\caption{
``Internal energies" $\langle A^{\rm 2D} \rangle,
\langle A^{\rm 2D}_{\rm s} \rangle, \langle A^{\rm 2D}_{\rm h} \rangle$ vs
$1/c_1$ for fixed $s_3/c_1=0.83$. We also show $C^{\rm 2D}$.
The intersection point of the two phase transition lines
(merging point of the AF and BEC phase transitions) is
marked  by an oval. At this region, 
$C^{\rm 2D}$
exhibits no anomalous behavior. Slightly anomalous behavior of 
$\langle A^{\rm 2D}_{\rm s} \rangle$ and 
$\langle A^{\rm 2D}_h \rangle$ cancel with each other and the
total $\langle A^{\rm 2D} \rangle$ is a regular function of $1/c_1$.
}
\label{energ}
\end{figure}

We first study the ``specific heat" (fluctuation of the action $A^{\rm 2D}$),
\be
C^{\rm 2D}=\frac{1}{V}\langle (A^{\rm 2D}-\langle A^{\rm 2D} \rangle)^2\rangle.
\label{sh}
\ee
In Fig.\ref{specificheats} we present $C^{\rm 2D}$ 
as a function of $s_3$ for various values of $c_1$.
We also present the ``specific heats" $C^{\rm 2D}_{\rm s}$ 
and $C^{\rm 2D}_{\rm h}$ of each term
$A^{\rm 2D}_{\rm s}$, $A^{\rm 2D}_{\rm h}$ in the action
as defined in (\ref{sh}).
As in the previous 3D finite-$T$ case, these 
results indicate that
there exist two phase transition lines in the $s_3-c_1$ plane.
They intersect at $c_1\simeq 5.0, s_3 \simeq 4.0$.
At the BEC transition points at high $c_1$'s (low $1/c_1$)
(see Fig.\ref{specificheats} for $c_1$=8.0 near $s_3\sim $7.0), 
$C^{\rm 2D}$ 
exhibits large values and large system-size dependence.

In order to see what happens at this intersecting point of the two 
phase transition lines, $c_1\simeq 5.0, s_3\simeq 4.0$,
we measured ``internal energies" $\langle A^{\rm 2D} \rangle,
\langle A^{\rm 2D}_{\rm s} \rangle, \langle A^{\rm 2D}_{\rm h} \rangle$,
and $C^{\rm 2D}, \; C^{\rm 2D}_{\rm s}, \; C^{\rm 2D}_{\rm h}$ as a function of $1/c_1$ 
along $s_3/c_1=0.83$.
We show the result in Fig.\ref{energ}.
At the intersection point, the total specific heat $C^{\rm 2D}$
exhibits no anomalous behavior, though $C^{\rm 2D}_{\rm s}$ 
and $C^{\rm 2D}_{\rm h}$ show peaks
at that point.
The calculations shown in Fig.\ref{energ} show an interesting 
phenomenon that anomalous behavior in $\langle A^{\rm 2D}_{\rm s} \rangle$ and 
$\langle A^{\rm 2D}_{\rm h} \rangle$ cancel with each other and the
total $\langle A^{\rm 2D} \rangle$ is a regular function of $1/c_1$.

\begin{figure}[b]
\begin{center}
\includegraphics[width=7.5cm]{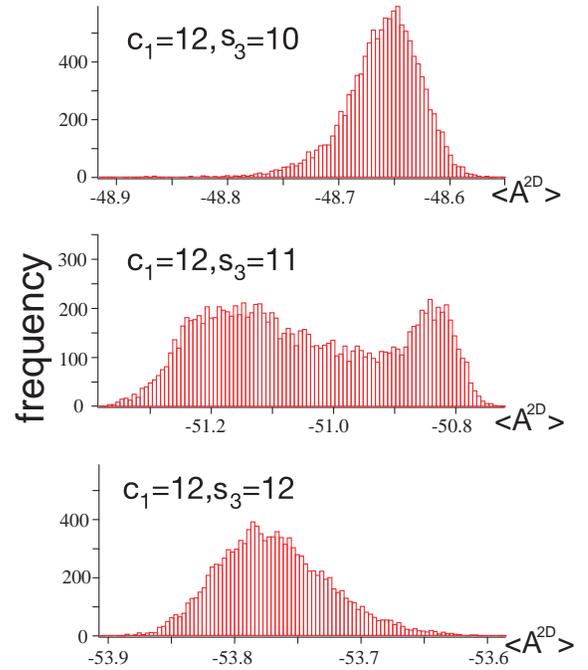}
\end{center}
\caption{
Distribution of $A^{\rm 2D}$. At $c_1=12, \; s_3=11$, it exhibits
a double-peak structure, 
whereas the other two do not. This result means
that the phase transition here is of first order.
}
\label{1-st-hist}
\end{figure}

To verify the order of the phase transitions, we measured
distribution of $A^{\rm 2D}$ for configurations generated 
through the MC steps.
In Fig.\ref{1-st-hist} we present
the distributions of $A^{\rm 2D}$ around $c_1=12, \; s_3=10,11,12$.
The distribution at the middle point, $s_3= 11$, shows a 
double-peak structure, so we concluded that the 
phase transition at this point is of first order.
We found that the phase transition separating the
 AF and AF+BEC phases is of first order,
 while the other three transition lines 
 are of second order.
 The difference of the order (1st vs. 2nd)
in the 3D and 2D models may be attributed to the difference
of holon hopping along the 3rd direction explained
at the end of Sect.4. The present 2D model has a $\bar{\phi}\phi$
uniform coupling, which put more weight for ordering of holons and
spinons than the 3D  model. 
This may generate the first-order transition in certain regions.

\subsection{Spin correlations}


\begin{figure}[b]
\begin{center}
\includegraphics[width=8cm]{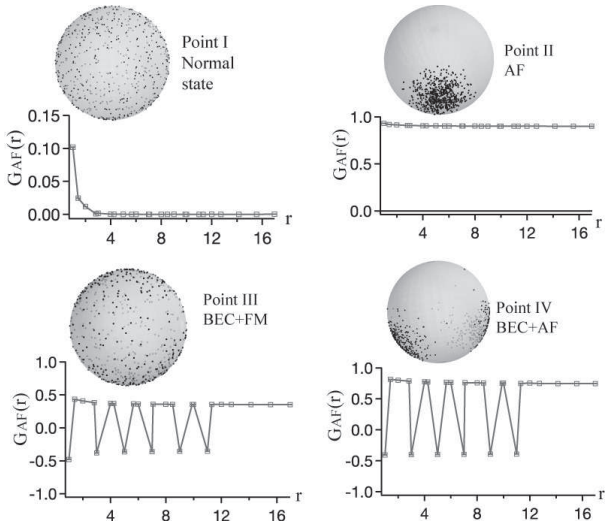}%
\end{center}
\caption{
Spin correlation function $G_{\rm AF}(r)$ of (\ref{gr2d})
and  spin snapshot at the points (I-IV) in Fig.\ref{Fig:phase_T=0}.
(I) $c_1=2, \; s_3=1$.
There is no long-range order of spins.
(II) $c_1=20, \; s_3=1$.
Endpoints of spins are centered near the south pole of the sphere,
indicating the (AF) long-range order.
(III) $c_1=4, \; s_3=6$.
Results indicate the alternative order, i.e., the FM order.
(IV) $c_1=20, \; s_3=30$.
Endpoints form two groups on the sphere, indicating
a canting ``order" of spins\cite{mft}.
}
\label{spin2-1}
\end{figure}

In this subsection, we investigate the spin correlation function
at equal time,
\be
G_{\rm AF}(r)\equiv \frac{1}{2V}\sum_{x,i}
\la \vec{S}_{x+ri}\vec{S}_{x} \ra,
\label{gr2d}
\ee
and show the snapshots of spin configurations.
In Fig.\ref{spin2-1} we present $G_{\rm AF}(r)$ and
the snapshots of spins 
at the typical four points (I-IV) in the phase
diagrams of Fig.\ref{Fig:phase_T=0}.
In the snapshots,  each $\vec{S}_x$ starts 
from the center of the unit sphere
and ends at a dot on the sphere.\\
(I) $c_1=2, s_3=1:$ Directions of spins are random, and
there is no long-range magnetic order.\\
(II) $c_1=20, s_3=1:$ This point in the phase diagram  
is located in the deep AF region.
It is obvious that there exists the AF 
long range order.\\
(III) $c_1=4, s_3=6:$ There appears the oscillative
 order around $G_{\rm AF}=0$, i.e., the  FM long range order 
 instead of the AF order.\\
(IV) $c_1=20, s_3=30:$
The even and odd site spins have their own magnetizations, 
$\vec{M}_{\rm e}$ and $\vec{M}_{\rm o}$, and these even and odd
magnetizations cant with each other.
This corresponds to the canting state studied in Ref.\cite{mft}.
In other word, there appear a component of a
 FM order in the background of AF long-range order.\\

\subsection{Gauge dynamics and topological objects}

\begin{figure}[b]
\begin{center}
\includegraphics[width=7.3cm]{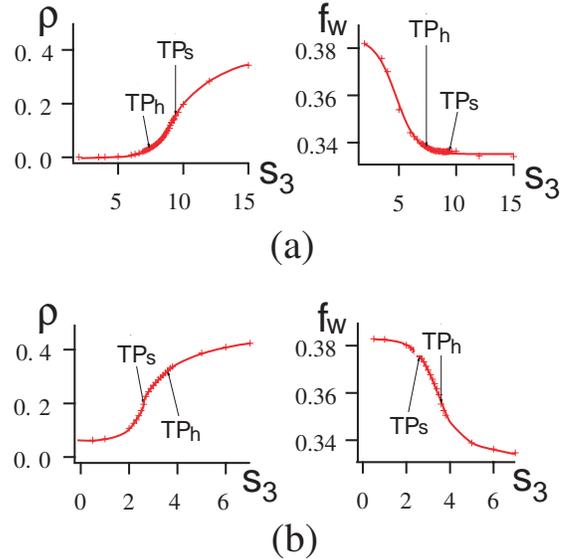}
\end{center}
\caption{
The instanton density $\rho$ and the flux density $f_W$ 
of (\ref{flux}) for 
(a) $c_1=8.0$
and (b) $c_1=4.0$.
``TP$_{\rm s}$"(``TP$_{\rm h}$") indicates the location of
the AF (BEC) phase transition point.
$\rho$  
changes  suddenly at TP$_{\rm s}$.
while $f_W$ changes  suddenly at TP$_{\rm h}$ as expected.
}
\label{inst}
\end{figure}

In this subsection, we study the gauge dynamics 
associated with two U(1) gauge fields; the spin hopping 
amplitude $U_{x\mu}$ and the dual U(1) holon hopping
amplitude $W_{xi}$ defined as follows;
\be
W_{xi}\equiv \frac{\tilde{z}_x \bar{z}_{x+i}}{|\tilde{z}_x \bar{z}_{x+i}|}
\in U(1).
\ee
For $U_{x\mu}$, we calculated the instanton density $\rho$ as in 
Sec.3\cite{instanton}, whereas for $W_{xi}$, we calculated its
gauge-invariant flux density, 
\be
f_W = \frac{1}{2\pi}\left<{\rm mod}\left[ 
-i\log(\bar{W}_{x2}\bar{W}_{x+2,1}W_{x+1,2}W_{x1}), 2\pi\right]\right>,
\nn
\label{flux}
\ee
as $W_{xi}$ has only the spatial components. 
In Fig.\ref{inst} we present  
$\rho$ and $f_W$ as functions of $s_3$;  (a) $c_1=8.0$
and (b) $c_1=4.0$.
For the both cases, it is obvious that
 $\rho$ and $f_W$ change their
values suddenly  at the relevant phase transition
points determined by $C^{\rm 2D}$.
We also verified that the singular behavior of the ``specific heat"
$C_{\rm s}\; (C_{\rm h})$ 
correlates to that of $\rho \; (f_W)$ as it is
expected.

$\rho$ influences the low-energy excitations of
the spinon sector, whereas  $f_W$ is related to the
holon hopping.
The large-$\rho$ region corresponds to the confinement phase
of spinons and the low-energy excitations there 
are described by the spin-triplet
$\bar{z}_x\vec{\sigma}z_x$.
On the other hand, in the small-$\rho$
 region of the AF state,
deconfinement of quanta $z_x$ takes place,  
and the low-energy excitations are
gapless spin waves.

Large fluctuations of $W_{xi}$ hinder coherent hopping of holons,
and induce large fluctuations of the holon field $\phi_x$.
However, as the holon density $\delta$ is increased, 
the holon hopping term stabilizes $W_{xi}$ and reduces $f_W$.
From our study of $\rho$ and $f_W$,  there seem to be some
correlations between them, the origin of which is of course 
the term $A^{\rm 2D}_{\rm h}$.

\subsection{Electron correlations}

In this subsection, we study the correlation functions
of single electrons and electron pairs 
in the 2D lattice in the various phases.
We define the bosonic electron operator, ${\cal B}_{x\sigma}$, 
as follows;
\begin{equation}
{\cal B}_{x\sigma}=
\phi^\dag_x\times\left\{
\begin{array}{rl}
z_{x\sigma},& \;\; {\rm for}\ \epsilon^{\rm 2D}_x=1, \\
\tilde{z}_{x\sigma},& \;\; {\rm for}\ \epsilon^{\rm 2D}_x=-1.
\end{array}
\right.
\label{Bx}
\end{equation}
${\cal B}_{x\sigma}$ is invariant
under the local U(1) gauge transformation.
We define also operators of
the spin-singlet and the spin-triplet  pairs  
of bosonic electrons on the NN sites at equal time,
\begin{eqnarray}
&& \Delta^{\rm s}_{xi}=
{\cal B}^\dagger_{x+i,\uparrow}{\cal B}^\dagger_{x\downarrow}
-{\cal B}^\dagger_{x+i,\downarrow}{\cal B}^\dagger_{x\uparrow},
\nonumber \\
&& \Delta^{\rm t}_{xi}=
{\cal B}^\dagger_{x+i,\uparrow}B^\dagger_{x\uparrow}
+B^\dagger_{x+i,\downarrow}B^\dagger_{x\downarrow}.
\label{CsCt}
\label{Deltax}
\end{eqnarray}
Then we introduce correlation functions  
of ${\cal B}_{x\sigma}$ and $\Delta^{\rm s,t}_{xi}$ 
at equal time as
\be
G_B(r)&\equiv&\frac{1}{4V}\sum_{x,i,\sigma}\la
{\cal B}^\dag_{x+ri,\sigma} {\cal B}_{x\sigma}\ra,\nn
G^{\rm s,t}(r) &\equiv& \frac{1}{2V}
\sum_{x,i\neq j}\la \bar{\Delta}^{\rm s,t}_{x+ri,j}
\Delta^{\rm s,t}_{xj}\ra.
\label{epc}
\ee

\begin{figure}[t]
\vspace{-0.5cm}
\hspace{-0.5cm}
\begin{center}
\includegraphics[width=9cm]{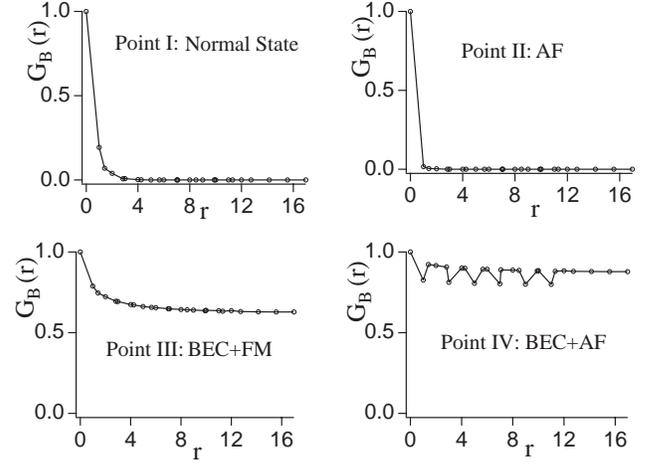}\\
\hspace{0.5cm}
\vspace{0.5cm}
(a) Correlations of single electrons \\
\hspace{-0.5cm}
\includegraphics[width=8.5cm]{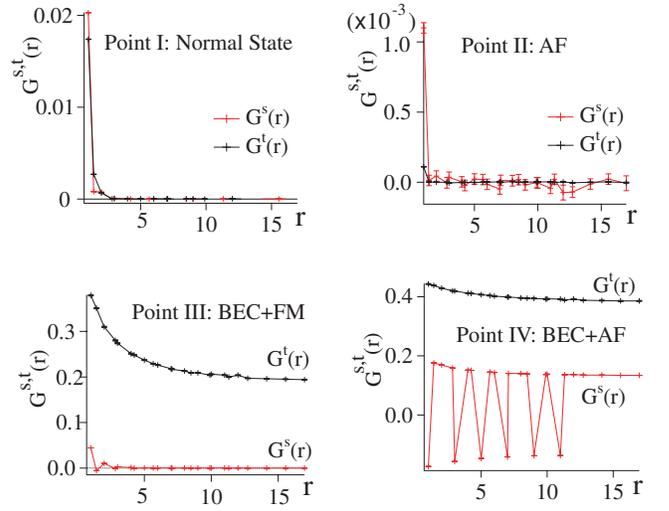}\\
(b) Correlations of electron pairs  \\
\end{center}
\caption{
Electron correlation functions of (\ref{epc})
at the four points (I-IV) of Fig.\ref{Fig:phase_T=0}: 
(a) $G_B(r)$ of single bosonic electrons;
(b) $G^{\rm s,t}(r)$ of spin-singlet 
and triplet pairs. 
Their meaning in each point is discussed in the text.
}
\label{cpcorr}
\end{figure}

\noindent
Their behaviors in the previous four points (I-IV) in  
 Fig.\ref{Fig:phase_T=0} are shown in Fig.\ref{cpcorr}.
Each point has the following properties; \\
 (I) NS phase and (II) AF phase:
 All the three functions,  $G_B(r), G^{\rm t,s}(r)$ 
 have no long-range correlations.\\
(III) BEC+FM phase:  
$G_B(r)$ and the triplet pair $G^{\rm t}(r)$ 
exhibit the long-range order, and 
the BEC takes place in these channels.\\
(IV) BEC+AF phase: all the three
functions exhibit the long-range order,
and the BEC takes place.\\
As discussed in Sect.3, these results are consistent
with the expectation that the two BECs, one of single electrons
and the other of (triplet) electron pairs, take place at the same time. 
Furthermore, these results and the previous results on the 
(AF and FM) spin orders indicate
that {\em the BEC order and spin order can be superimposed}
in certain region.
This is another example of the ``spin-charge" separation.


\section{Conclusion}
\setcounter{equation}{0} 

In this paper, we have investigated the bosonic $t$-$J$ model
in the CP$^1$-spinon  + U(1) Higgs-holon representation. 
This model is introduced by replacing the fermionic holons with the
bosonic ones in the slave-fermion $t$-$J$ model.
We studied its phase structure and the dynamical properties  
both in the 3D finite-$T$ model and the 2D $T=0$ model.
In particular, we are interested in the interplay of the BEC and 
AF order, because a coherent holon hopping is required for the BEC,
whereas it hinders the AF long-range order.
Our study by means of the MC simulations exhibits 
a phase diagram
in which  the coexistence region of AF and BEC orders appears.
This result suggests that in the fermionic $t$-$J$ model 
a MIT
takes place at a finite hole concentration $\delta_c$
and that phase transition point is located within the AF region of the 
spin dynamics.
From the result of the present paper, we also expect that
as $\delta$ is increased further, the BEC phase
transition through the formation of electron pairs 
and their BEC takes place in the AF region.
Actually, this problem can be studied 
 in the framework of  the fermionic $t$-$J$ model 
in the slave-fermion representation.
Explicitly, one may obtain the effective model 
by integrating out the fermionic holon field
by the hopping expansion.
The resultant model includes only the bosonic variables;
$z_x$ of spinons, ``order parameter" of the coherent hopping 
of holons (i.e., the dual gauge field $W_{xi}$),
 and the SC order parameter
of electron pairs, which can be analyzed numerically\cite{btJ2}.
For this scenario one should include
some interaction terms that we have ignored in Sect.4
in our action, but  have been generated in the process of
integrating half of the spinon variables\cite{CP1}.
Among these terms, there is an attractive interaction 
between fermionic holons at NN sites\cite{T0}. 
This term favors formation of holon pairs, and their
coherent condensation gives rise to the superconductivity.

In an optical lattice, cold atoms at each well can have (pseudo-)spin 
degrees of freedom $2s+1=1,2,\cdots$, and there appears an AF 
interaction among them as a result of the repulsion.
In this case, the CP$^1$ constraint is changed to 
$\sum_{a=1}^2|z_{xa}|^2=2s$,
but this change of the normalization of $z_{x}$ can be absorbed by
the parameter $c_1$.
Then cold atom systems with {\em one particle per well} can be described by 
the CP$^1$ model like $A_{\rm s}$ of 
Eq.(\ref{action})\cite{so5}.

We expect that 
$A^{\rm 2D}$ of (\ref{action2d}) 
describes a 2D system of 
{\it bosonic} cold atoms with (pseudo-)spin 
and {\em  repulsive interaction} in an
optical lattice {\em slightly away from the case of one-particle per well}.
For such system of cold atoms 
the phase diagram, Fig.\ref{Fig:phase_T=0}, indicates 
the coexistence of a BEC   and
a long-range order of an internal symmetry \cite{cold}.

Let us comment on   the effect of our replacement 
(\ref{replace}) of fermionic holons by
Higgs field with a uniform amplitude,
$\sqrt{\mathstrut \delta}\phi_x$.
Being compared with the faithful bosonic model of the
original  $t$-$J$ model, 
this replacement certainly ignores fluctuations of
the amplitude (density) of holon variable, which disfavors the BEC
of holons. However, we expect that  the  BEC we obtained
in the present model should survive in the faithful 
bosonic $t$-$J$ model, although location of the transition 
curves may change.
This problem is under study and we hope that result will be
reported in a future publication\cite{NSIM}.

\bigskip
\acknowledgments
This work was partially supported by Grant-in-Aid
for Scientific Research from Japan Society for the 
Promotion of Science under Grant No.20540264.



\begin{references}
\bibitem{hcb}
The derivation of the present model may have different routes.
For example, one may start from the $t$-$J$ model of electrons
and replace the electron operator $C_{x\sigma}$ 
by a canonical boson operator
$D_{x\sigma}$. Then the condition to exclude the 
 double-occupancy states may be respected by introducing 
 the slave-particle
representation as $D_{x\sigma}=\Phi^\dagger_x a_{x\sigma}$
where $\Phi_x$ is the hard-core-boson operator.
The substitution $\Phi_x \rightarrow \sqrt{\mathstrut \delta}\ \phi_x$
gives the present model.
\bibitem{irrelevant}
Precisely speaking, the action involves other terms that come
from the representation (\ref{a}).
Their explicit form and roles are
discussed in details in Ref.\cite{CP1}. 
We ignore them here for simplicity and
partly because the interplay
of AF and BEC can be studied without these terms.
\bibitem{su2}
In the model studied in Ref.\cite{btj}, 
the AF order at $T=0, \alpha=1, \delta=0$ is
suppressed for $\delta > 0$ even at $T=0, \alpha=1$.
The author of Ref.\cite{btj} explains it 
because the global SU(2) spin symmetry 
of the Hamiltonian at $\alpha=1, \delta=0$ is 
lost for $\delta > 0, \alpha=1$ 
(due to the omitted $m_i m_j+n_i n_j$ term). 
On the other hand, the action (\ref{action}) 
preserves the SU(2) symmetry for any $\delta$
as in the original $t$-$J$ model.
In fact, the spin rotation is induced by $z_{x}\rightarrow
g z_{x}$ where
$g \in$ SU(2). Then $\bar{z}_{x+\mu} z_x \rightarrow
\bar{z}_{x+\mu}\bar{g}g z_x,\ \tilde{z}_{x+\mu} \bar{z}_x
\rightarrow {\rm det}g\cdot \tilde{z}_{x+\mu} \bar{z}_x$,
both of which remain unchanged.
\bibitem{sraf}
A qualitative estimation of the relative
errors associated with this treatment
is given for $\delta=0$ as $O(\exp(-12\beta J))$
in Sec.2(B6) of the first paper of Ref.\cite{CP1}. 
We shall return this point 
at the end of Sec.5A after obtaining  
 the phase diagram, Fig.\ref{Fig:phase_T=0}.
\bibitem{acp1}
More precisely, the Berry-phase term should be added to $A_{\rm s}$ 
of (\ref{ACP1}).
This phase is known to suppress monopole (instanton) configurations
of the gauge field $U_{x\mu}$.
From the study on both the compact and noncompact U(1) gauge models
of CP$^1$ boson\cite{TIM1,noncompact}, we expect that the models 
with and without the Berry phase have a similar phase structure 
though the critical exponents, etc.,
have different values in two models.
\bibitem{eta}
In Ref.\cite{CP1} we 
write the action using $\eta_{\rm o}$ instead of 
$\psi_{\rm o}^{\rm new}$.
They are related as $\eta_{\rm o} = \bar{\psi}_{\rm o}^{\rm new}$.
\bibitem{c1s3}
 $c_1$ is estimated\cite{CP1} as
$c_1 =J a_3 = J a_3' a/c=a_3'/\sqrt{2}$
where $a_\mu$ is the lattice spacing in the $x_\mu$ direction
($a\equiv a_1= a_2$),
$a_3'  \equiv (c/a)  a_3$, and 
$c\equiv\sqrt{2}Ja$ is the ``speed of light".
$a_3'$ is a dimensionless number of $O(1)$, 
the optimal value of which
is to be determined by renormalization-group argument. 
The extra factor $1/J$ in $s_3$ comes from the
rescaling $a_3 \rightarrow a_3'$.
\bibitem{mc3d}
For MC simulations of the present (2+1)D system,
we used the similar set up and procedures used in Sect.3; 
i.e., $V=L^3$ up to $L=30$, etc.
\bibitem{instanton}
We use ``instanton" here instead of
``monopole" because the present system is defined in $2+1$ dimensions.
\bibitem{T0} This term was called $T_0$ in Eq.(2.39) of 
the first paper in Ref.\cite{CP1}.
\end{references}
\end{document}